\documentclass[final,5p,times,twocolumn]{elsarticle}

\usepackage{amssymb}

\usepackage{amsmath}
\usepackage{amssymb}
\usepackage{soul}
\usepackage{enumerate}

\usepackage{helvet}
\usepackage{courier}
\usepackage{amsthm}
\usepackage[T1]{fontenc}
\usepackage[latin9]{inputenc}
\usepackage{makeidx}         
\usepackage{graphicx}        
\usepackage[bottom]{footmisc}

\usepackage{hyperref}
\hypersetup{colorlinks=true,
  linkcolor=blue,
  anchorcolor=blue,
  citecolor=red,
  urlcolor=brown,
  pdfauthor={Adriano Barra and Antonio Moro}}

\providecommand{\be}{\begin{equation}}
  \providecommand{\ee}{\end{equation}}
\providecommand{\bea}{\begin{eqnarray}}
  \providecommand{\eea}{\end{eqnarray}}
\providecommand{\beas}{\begin{eqnarray*}}
  \providecommand{\eeas}{\end{eqnarray*}}

\providecommand{\beni}{\begin{equation*}}
  \providecommand{\eeni}{\end{equation*}}

\providecommand{\bw}{\begin{widetext}}
  \providecommand{\ew}{\end{widetext}}

\newcommand{\benumerate}{\begin{enumerate}}
\newcommand{\eenumerate}{\end{enumerate}}

\newcommand{\vep}{\varepsilon}
\newcommand{\der}[2]{\frac{\partial #1}{\partial #2}}

\newcommand{\dersec}[2]{\frac{\partial^{2} #1}{\partial #2^{2}}}

\newcommand{\dermixd}[3]{\frac{\partial^{2} #1}{\partial #2 \partial #3}}

\journal{Physica D}

\begin{document}

\begin{frontmatter}



\title{Integrable extended van der Waals model}


\author[label1,label2]{Francesco Giglio}
\author[label2,label3]{Giulio Landolfi}
\author[label1]{Antonio Moro\cortext[cor1]{contact: antonio.moro@northumbria.ac.uk}}

\address[label1]{Department of Mathematics and Information Sciences, University of Northumbria at Newcastle upon Tyne, Camden St., NE21XE, UK}

\address[label2]{Department of Mathematics and Physics {\it "Ennio de Giorgi"} of the University of Salento, I-73100 Lecce, Italy}

\address[label3]{I.N.F.N. Sezione di Lecce, via Arnesano I-73100 Lecce, Italy}

\begin{abstract}
Inspired by the recent developments in the study of the thermodynamics of van der Waals fluids via  the theory of nonlinear conservation laws and the description of phase transitions in terms of classical (dissipative) shock waves, we propose a novel approach to the construction of multi-parameter generalisations of the van der Waals model. The theory of integrable nonlinear conservation laws still represents the inspiring framework. Starting from a macroscopic approach, a four parameter family of integrable extended van der Waals models is indeed constructed in such a way that the equation of state is a solution to an  integrable nonlinear conservation law linearisable by a Cole-Hopf transformation. This family is further specified by the request that, in regime of high temperature, far from the critical region, the extended model reproduces asymptotically the standard van der Waals equation of state. We provide a detailed comparison of  our extended model with two notable empirical models such as Peng-Robinson and Soave's modification of the Redlich-Kwong equations of state. We show that our extended van der Waals equation of state is compatible with both empirical models for a suitable choice of the free parameters and can be viewed as a {\it master} interpolating equation. The present approach also suggests that further generalisations can be obtained by including the class of dispersive and viscous-dispersive nonlinear conservation laws and could lead to a new type of thermodynamic phase transitions associated to nonclassical and dispersive shock waves.
\end{abstract}

\begin{keyword}
van der Waals model \sep nonlinear conservation laws \sep integrability
\PACS 05.70.Ce \sep 64.10.+h \sep 02.70.-c \sep 64.60.Bd \sep 64.60.fd 

\MSC[2010] 82B26 \sep 35L67 \sep 35Q79


\end{keyword}

\end{frontmatter}


\section{Introduction}
The van der Waals model first introduced to describe liquid-vapour coexistence in simple fluids \cite{refvdW} is now considered a classical paradigm for the description of phase transitions for a large family of physical systems (see e.g. \cite{Callen, Stanley}.) In fact, the celebrated van der Waals equation of state
\begin{equation}
\label{vdweq}
\left (P +\frac{a}{v^{2}} \right ) (v -  b) = R T
\end{equation}
can be both obtained from first principles as a mean field approximation for a system of hard core particles with electrostatic interaction  and also via a simple and intuitive heuristic derivation (see e.g.~\cite{Stanley}). An alternative mean field approach has been recently introduced in~\cite{BM} that allows to rigorously establish a formal analogy between the van der Waals model and magnetic mean field models such as the Curie-Weiss model and its multi-component extensions~\cite{BZ,CW,sumrule,Genovese}. Interestingly, the phenomenological approach has played a key role over the decades, in particular for chemical engineering applications~\cite{Soave,RK,PR,pitzer, SIP} aimed at providing a more accurate description of composite systems, such as solutions and multi-phase systems, for which a statistical physics approach and a mean field theory is not currently available. It should also be mentioned that although the van der Waals models catches many fundamental qualitative features of phase transitions in fluids, a quantitative demands for more accurate, though empirical, equations of state. 

Hence, a number of alternative and/or generalised approaches to the van der Waals theory have been introduced based on both mean field theory or phenomenological approaches (see e.g~\cite{Landau, Toledano,Widom,WR, Lebowitz, Kampen, Griffiths, Soave,RK,PR}).

More recently, in a series of papers~\cite{BM,Moro,DM} it was observed that the van der Waals equation of state can be interpreted as a particular solution to an integrable nonlinear partial differential equation (PDE), that is equivalent to the first law of thermodynamics, specified by assigning a particular isothermal/isobaric curve, possibly far from the critical region.
The choice of such particular isothermal/isobaric curves is equivalent to the choice of an initial datum for the PDE and it is sufficient to fix uniquely the solution. Therefore, initial data can be used to  parametrise a large family of integrable models of which the van der Waals equation is one example. Moreover, as shown in~\cite{Moro} and \cite{BM} this approach formulated through integrable nonlinear conservation laws can be extended to the critical region relying on an asymptotic procedure that automatically encodes Maxwell's equal areas rule and provides an asymptotic analytic description of phase transitions in terms of shock wave solutions to hyperbolic nonlinear conservation laws~\cite{Whitham}.

We note that the correspondence between phase transitions and shock solutions of nonlinear PDEs has been also observed and studied in depth in the context of mean field spin models~\cite{BSZ,Tantari,BDGM,BZ,CW,sumrule,Genovese}, chemical kinetics and cybernetics~\cite{Agliari}, neural networks~\cite{Agliari2} and their statistical mechanical description.

In the present work we propose a new method to extend a thermodynamic model by the request that the equation of state remains a solution to a suitable extended nonlinear conservation law. Although the approach can be easily formulated for any model within the general class studied in  \cite{DM}, for the sake of simplicity we will focus on the construction of the extension to the van der Waals model only. Our fundamental assumptions are:
\begin{enumerate}[i)]
\item 
\label{prop1} the van der Waals model is assumed to be accurate in regime of high temperature and low density, so that the proposed extension must asymptotically reproduce the van der Waals equation of state;
\item \label{prop2} the nonlinear conservation law is required to be $C-${\it integrable}, that is linearisable via a Cole-Hopf transformation.
\end{enumerate}
Property (\ref{prop1}) relies on the idea that if the gas is in thermodynamic equilibrium and  sufficiently rarified, particles can be modelled as rigid spheres interacting by a Coulomb potential. In this regime, the van der Waals model is expected to be sufficiently accurate. Property (\ref{prop2}) is based on the result obtained in \cite{BM} where it was shown that the van der Waals mean field model is completely integrable by linearisation. In fact, it was proven that volume density fulfils a nonlinear PDE, in the class of conservation laws, that is linearisable to the Klein-Gordon equation via a Cole-Hopf transformation. Hence, we require that the extended model preserves this property. The family of models so obtained is parametrised by four arbitrary constants and contains the van der Waals model as a particular case.
These models can be viewed as a two parameter deformation of the van der Waals model. The thermodynamic limit is then calculated via a standard asymptotic expansion in the small expansion parameter $\eta = 1/N_{A}$ where $N_{A}$ is Avogadro's number the associated phase diagrams are then evaluated. We also observe that the Cole-Hopf transformation provides the natural extension of the mean field van der Waals partition function derived in~\cite{BM}.

We finally compare our model with two well-known phenomenological extensions of the van der Waals model: the Soave's modification of the Redlich-Kwong (SRK) equation of state~\cite{Soave,RK} and the Peng and Robinson (PR) equation of state~\cite{PR}. We show that our model exactly reproduces both Soave-Redlich-Kwong and Peng-Robison critical points for a suitable choice of the parameters and can be proposed as the interpolating model. Our analysis suggests that the proposed model is suitable for describing a wide class of real systems such as, solutions and multi-component thermodynamic systems, within the range of applications of both SRK and PR equations of state.

The paper is organised as follows:  In Section~\ref{sezione 2} we introduce the general macroscopic  model and formulate of the first principle of thermodynamics  in terms of a nonlinear PDE. The integrability condition by linearisation via a Cole-Hopf transformation, given a suitable expansion in the order parameter, allows us to specify the class of models such that the equation of state is obtained as a solution to an integrable nonlinear PDE. In Section \ref{sezione 3} we focus on the sub-family of models that can be viewed as an extension of the van der Waals model, provide their full characterisation and construct a natural extension of the mean field partition function. We evaluate the critical asymptotics and phase diagrams in Section \ref{sezione 4}. Section \ref{sezione 5} is devoted to a detailed comparison of our model with PR and SRK models.
Final remarks and an outlook on future works is included in Section~\ref{sezione 6} where we also argue that the present method can be applied to construct a wider class of models where the equation of state is obtained as a solution to a dissipative and dispersive equation implying a richer critical phenomenology.

\section{General model equations}
\label{sezione 2}
Let us consider $n$ moles of a gas whose physical state is determined by its volume $V$, pressure $P$ and temperature $T$. The number of particles is $N = n N_{A}$ where $N_{A}$ is Avogadro's number.
Introducing the Gibbs thermodynamic potential $G = E - TS + PV$ where $E$ is the internal energy, the first principle of thermodynamic reads as
\begin{equation}
\label{firstlaw}
d G = - S dT + V dP,
\end{equation}
where the $S$ is the entropy of the system.
Introducing the  variables 
\begin{equation}
x =\frac{P}{T}   \qquad t = \frac{1}{T} 
\end{equation}
and the molar volume density 
\[
v = \frac{V}{n}
\]
the balance equation~(\ref{firstlaw}) is written as follows
\begin{equation}
\label{firstlaw2}
d \psi = \varepsilon dt + v dx
\end{equation}
where $\psi := t G/n$  and $\varepsilon := E/n$ is the internal energy per mole unit. We also observe that the Equation (\ref{firstlaw2}) is locally equivalent to the closure condition
\begin{equation}
\label{balance}
\der{v}{t} = \der{\varepsilon}{x},
\end{equation}
also known as Maxwell relation~\cite{Callen}.
In the present paper, we focus on the class of models such that the internal energy function admits the following expansion in terms of the small parameter $\eta = 1/N_{A}$
\begin{equation}
\label{Eexp}
\vep = \vep_{0}(v) + \eta \vep_{1}(v) \der{v}{x} + \eta \vep_{2}(v) \der{v}{t} + O(\eta^{2}) + g(t)
\end{equation}
where $\vep_{i}(v)$, $i=0,1,2$ are functions of the one single variable $v$ and $g(t)$ is an arbitrary function of its argument $t$.  A similar expansion has been considered for the entropy function in 
\cite{Moro}. We point out that the internal energy for the  van der Waals model is a separable function of volume and temperature (see e.g.~\cite{Landau})
\[
E_{vdW} = \frac{N a^{2}}{V} + g(T).
\]
Moreover, perturbative approaches based on a double parameter expansion of Lennard-Jones potentials (see~\cite{Baker} and also~\cite{Mansoori} for a review) lead to equations of states which explicitly depend on the compressibility and the coefficient of thermal expansion, and then on derivatives of the volume with respect to pressure and temperature respectively. The ansatz~(\ref{Eexp}) reflects the above two properties. In the region of thermodynamic variables, $x$ and $t$, where the derivatives of the molar volume are bounded, $O(\eta)$ terms can be neglected  and we can approximate
\[
\vep  \simeq  \vep_{0}(v) + g(t).
\]
Equation~(\ref{balance}) gives the Riemann-Hopf type equation of the form
\begin{equation}
\label{RH}
\der{v}{t} = \vep_{0}'(v) \der{v}{x} 
\end{equation}
with the notation $\vep_{0}'(v) = d \vep_{0}/dv$. The class of van der Waals type thermodynamic systems described by the equation~(\ref{RH}) admits the following equation of state
\begin{equation}
\label{eqstate_gen}
x + \vep_{0}'(v) t = f(v)
\end{equation}
where $f(v)$ is an arbitrary function of its argument and parametrises the family of models associated to the equation~(\ref{RH}). The equation~(\ref{eqstate_gen}) defines implicitly the general solution $v=v(x,t)$ to the Riemann-Hopf type equation~(\ref{RH}). For the special choice
\begin{equation}
\label{parf}
\vep_{0}(v) = -\frac{a}{v} \qquad f(v) = \frac{R}{v-b}
\end{equation}
this family of models provides just the van der Waals equation of state
\begin{equation}
\label{vdw_eqst}
\left (P +\frac{a}{v^{2}} \right ) (v -  b) = R T
\end{equation}
where $R\simeq 8.3145 \; J/mol K$ is the gas constant, $b$ is the molar hard core volume and $a$ is the macroscopic parameter describing the macroscopic effect of long range electrostatic interactions.
As it was pointed out in \cite{DM} equations of state of the form (\ref{eqstate_gen}) can be interpreted as a nonlinear wave solution to a system of hydrodynamic type. Such solutions generically break in finite time and the breaking point corresponds to the critical point associated to the gas-liquid phase transition.

Let us now consider the internal energy asymptotic expansion (\ref{Eexp}) up to $O(\eta^{2})$. In this case, the truncated equation~(\ref{balance}) reads as follows
\begin{equation}
\label{balance2}
\der{v}{t} = \der{}{x} \left[\vep_{0}(v) + \eta \vep_{1}(v) \der{v}{x} + \eta \vep_{2}(v) \der{v}{t} \right].
\end{equation}
We note that away from the critical region solutions to the equation~(\ref{balance2}) are expected to converge, at the leading order, to  the  solution (\ref{eqstate_gen}) of the Rieman-Hopf equation~(\ref{RH}) discussed above. However, near the critical point, where derivatives of the solutions of the van der Waals equation diverge, the $O(\eta)$ terms will importantly affect the leading order,  providing a result that is consistent with experimental observations also within the critical region.
Introducing the function $\varphi$ via the Cole-Hopf transformation
\begin{equation}
\label{colehopf}
v(x,t) = \sigma \eta \frac{\partial \log \varphi}{\partial x} (x,t; \eta) \, ,
\end{equation}
where $\sigma$ is a non-vanishing real constant, Equation~(\ref{balance2}) transforms into the following nonlinear PDE
\begin{equation}
\label{phieq0}
\sigma^{2} \eta^{2} \left [\vep_{2} \dermixd{\varphi}{x}{t} + \vep_{1} \dersec{\varphi}{x} \right] - \eta (\sigma  \vep_{2} v +\sigma^2  ) \der{\varphi}{t} + (\sigma \vep_{0} - \vep_{1} v^{2}) \varphi = 0.
\end{equation}
We are interested in the class of nonlinear conservation laws of the form~(\ref{balance2}) that are linearisable via a Cole-Hopf transformation. We shall choose the functions $\vep_{0}(v)$, $\vep_{1}(v)$ and $\vep_{2}(v)$ in such a way that the equation~(\ref{phieq0}) reduces to a linear PDE. More precisely, we impose
\begin{align*}
&\sigma^2 \vep_{2} = c_{1}  B &  \sigma \vep_{2} v + \sigma^{2} = -c_{3}&  B \\
&\sigma^{2}  \vep_{1} = c_{2} B  & \sigma \vep_{0} -  \vep_{1} v^2 = c_{4}& B
\end{align*}
where $B = B(v)$ is a function to be determined and the $c_{i}$'s, are arbitrary real constants. Solving the above linear system for $\vep_{0}(v)$, $\vep_{1}(v)$ and $\vep_{2}(v)$ and $B(v)$ we obtain
\begin{equation}
\label{Eis}
\vep_{0} = - \frac{c_{2} v^{2} + c_{4} \sigma^{2}}{c_{1} v  + c_{3} \sigma} \qquad \vep_{1} = -\frac{c_{2} \sigma}{c_{1} v  + c_{3} \sigma} \qquad \vep_{2} = -\frac{c_{1} \sigma}{c_{1} v  + c_{3} \sigma}
\end{equation}
and
\[
B = -\frac{\sigma^{3}}{c_{1} v  + c_{3} \sigma}.
\]
Consequently, the function $\varphi$ satisfies the following  linear equation 
\begin{equation}
\label{phieq}
\eta^{2} \left (c_{1} \dermixd{\varphi}{x}{t} + c_{2} \dersec{\varphi}{x} \right) + \eta c_{3} \der{\varphi}{t} + c_{4} \varphi = 0.
\end{equation}
Equation~(\ref{phieq}) defines a four-parameter integrable extension of the standard van der Waals model and, for the choice of coefficients~(\ref{Eis}), it is equivalent to the equation (\ref{balance2}) on all solutions such that the Cole-Hopf transformation is defined. We will explicitly construct the solution to the Equation~(\ref{phieq}) at the leading order in the parameter $\eta$ and then the leading order solution to the equation~(\ref{balance2}). Let us observe that the particular case $c_{2} = c_{3} = 0$, where the equation (\ref{phieq}) reduces to the Klein-Gordon equation, has been obtained in \cite{BM} and the function $\varphi$ is interpreted as the mean field partition function. This solution, outside the critical region, and in the limit $\eta \to 0$, gives the van der Waals equation of state (\ref{vdw_eqst}) for the volume $v$ defined via~(\ref{colehopf}).

Finally, the asymptotic expansion of the internal energy up to $O(\eta^{2})$ is given by the following formula
\begin{equation}
\label{epsexp}
\vep \simeq -\frac{1}{c_{1} v + c_{3} \sigma} \left [ c_{2} v^{2} + c_{4} \sigma^{2} + \sigma \eta \,\left(c_{1} \der{v}{t} + c_{2}  \der{v}{x} \right)   \right ]  + g(t)
\end{equation}
and the equation~(\ref{balance2}) for the volume reads as follows
\begin{equation}
\label{balance3}
\der{v}{t} +\der{}{x} \left\{ \frac{1}{c_{1} v + c_{3} \sigma} \left[ c_{2} v^{2} + c_{4} \sigma^{2} + \sigma \eta \, \left(c_{1} \der{v}{t} + c_{2}  \der{v}{x} \right) \right ]\right\} = 0.
\end{equation}
Solutions $v= v(x,t;\eta)$ to the nonlinear conservation law~(\ref{balance3})  provide the equation of state for the family of models associated to the internal energy admitting the expansion the form~(\ref{epsexp}). We also observe that the integrable nonlinear PDEs~(\ref{balance3}) belongs to the more general family of integrable viscous conservation laws as classified in~\cite{ALM}. The integrability implies the existence of infinitely many commuting flows in the form of conservation law. In the thermodynamic context they involve the existence of infinitely many functions of state and corresponding conjugated variables. We also point out that the linearisable equation~(\ref{balance3}) can not be obtained as a particular case of the classification procedure discussed in~\cite{Svinolupov} as it is not of evolutionary type. \\
Equation~(\ref{balance3}) is indeed an example of integrable viscous conservation law that can be written in non-evolutionary but finite form. Based on the results in~\cite{ALM,ALM2} we can check that the equation~(\ref{balance3}) is a nonlocal symmetry of the Burgers equation. This can be straightfowardly done by computing the {\it viscous central invariant.
The notion of viscous central invariant has been introduced in~\cite{ALM} where it was conjectured that all scalar integrable viscous conservation laws are parametrised by one function of a single variable.  Indeed, observing that up to $O(\eta^{2})$ equation~(\ref{balance3}) can be written in evolutionary form as follows
\begin{equation}
\der{v}{t} = \der{}{x} \left [ f(v) + \eta A(v) \der{v}{x} + O\left (\eta^{2} \right ) \right]
\end{equation}
where
\begin{align*}
f(v) &= -\frac{c_2 v^2+c_4 \sigma^2}{c_1 v +c_3 \sigma } \\
A(v) &= \frac{\sigma }{c_1 v+ c_3 \sigma }  \left [-\frac{c_{1}^{2}  \left(c_2 v^2+ c_4 \sigma ^2\right)}{(c_1
v+c_3 \sigma )^2}+\frac{2 c_1 c_2 v}{c_1 v+c_3 \sigma} - c_2 \right ],
\end{align*}
the viscous central invariant $a(v)$ is calculated via the formula~\cite{ALM2}
\[
a(v) = \frac{2 A(v)}{f''(v)} 
\]
which gives $a(v) = \sigma$. A constant viscous central invariant proves that equation~(\ref{balance3}) belongs to the Burgers hierarchy. 
}
We finally observe that a new example of integrable viscous conservation law associated to a nonlinear (rational) viscous central invariant can be constructed via the hodograph transformation (interchange of dependent and independent variable) of the form
\[
\psi = \psi(x,t) \quad  \rightarrow \quad x = x(\psi,t)
\]
where $\psi$ is the potential such that $v =\partial \psi / \partial x $. Setting $u(\psi,t) = \partial x / \partial \psi$, equation~(\ref{balance3}) transforms to
\begin{equation}
\label{balance4}
\der{u}{t} = \der{}{\psi} \left\lbrace \left (u^{2} + \eta \vep_{2} \der{u}{\psi} \right)^{-1} \left [ -u^{3} \vep_{0} + \eta \left(\vep_{1} \der{u}{\psi} + u \vep_{2} \der{u}{t} \right) \right ]  \right \rbrace =0,
\end{equation}
where $\vep_{i} = \vep_{i}(v)$, $i=0,1,2$ are given in~(\ref{Eis}) and $v = \psi_{x} = 1/u$.
The viscous central invariant $a(u)$ of the equation~(\ref{balance4}) is the rational function
\begin{equation}
a(u)= - \left . \ \frac{2u}{\vep_{0}(v)''} \left(\vep_{1}(v) + \vep_{0}'(v) \vep_{2}(v) \right) \right |_{v = u^{-1}}
\end{equation}
The equation~(\ref{balance4}) is a first new example of integrable viscous conservation law with nonlinear viscous central invariant.
The  standard van der Waals case is obtained for the particular choice of the parameters $c_2=c_3=0$. This case is associated to the linear viscous central invariant $a(u)= - \sigma u$ that characterises the viscous analog of the Camassa-Holm hierarchy as studied in~\cite{ALM,Falqui}.

\section{Extended van der Waals model}
\label{sezione 3}
In this section we study the model equation~(\ref{phieq}) viewed as a deformation of the van der Waals model. It was shown in~\cite{BM} the van der Waals mean field partition function satisfies  Klein-Gordon equation, that is obtained as a particular case of~(\ref{phieq}) for $c_{2} = c_{3} = 0$ with $c_{1} \neq 0$ and $c_{4} \neq 0$.  Hence, dividing the equation~(\ref{phieq}) by $c_{1}$, that is assumed to be non zero, we have
\begin{equation}
\label{phieq2}
\eta^{2} \left ( \dermixd{\varphi}{x}{t} + r_{2} \dersec{\varphi}{x} \right) + \eta r_{3} \der{\varphi}{t} + r_{4} \varphi = 0.
\end{equation}
where we have introduced the parameters
\[
r_{2} = \frac{c_{2}}{c_{1}} \qquad r_{3} = \frac{c_{3}}{c_{1}} \qquad r_{4} = \frac{c_{4}}{c_{1}}.
\]
Hence, the expansion coefficients for the internal energy~(\ref{epsexp}) read as follows
\[
\vep_{0} = -\frac{r_{2} v^{2} + r_{4} \sigma^{2}}{v + r_{3} \sigma} \qquad \vep_{1} = \frac{r_{2} \sigma}{v + r_{3} \sigma} \qquad \vep_{2} = - \frac{\sigma}{v + r_{3} \sigma}.
\]
The request that the function $\vep_{0}$ reduces to the internal energy expansion coefficient for the van der Waals model~(\ref{parf}) immediately implies that
\[
r_{4} =\frac{a}{\sigma^{2}},
\]
so that
\begin{equation}
\label{eps0spec}
\vep_{0} = -\frac{r_{2} v^{2} + a}{v + r_{3} \sigma}.
\end{equation}
The extended model so constructed can consequently be viewed as two-parameter (i.e. $r_{2}$ and $r_{3}$) family deformation of the standard van der Waals model. The corresponding equation of state~(\ref{eqstate_gen}) is given by
\begin{equation}
\label{eqstate_part}
x  + \left [ \frac{r_{2} v^{2} + a}{(v- R r_{3})^{2}} - \frac{2 r_{2} v}{v- R r_{3}} \right] t - \frac{R}{v- b} = 0.
\end{equation}
One can immediately check that for $r_{2}=r_{3} = 0$, the equation~(\ref{eqstate_part}) reduces to the van der Waals equation~(\ref{vdw_eqst}). Similarly to the van der Waals equation, the integrable extended van der Walls (IEW) equation of state~(\ref{eqstate_part}) provides a description of the thermodynamic system in the case where the internal energy expansion in the parameter $\eta$ is truncated at the leading order and, as well known, it has local validity only, outside the region of thermodynamic variables where the system undergoes a phase transition.

Based on the considerations in~\cite{BM}, a global asymptotic solution can be constructed by solving the equation~(\ref{phieq2}) obtained from the first order expansion of the internal energy~(\ref{epsexp}). 

Let us consider the class of solutions to the equation~(\ref{phieq2}) of the form
\begin{equation}
\label{formalsol}
\varphi = \int_{b}^{\infty} \; \exp \left [{\frac{x v + \vep_{0}(v) t - \tilde{g}(v)}{\sigma \eta}} \right ] \; dv,
\end{equation}
where $\tilde{g}(v)$ is an arbitrary function of its argument.
At this stage, the above solution~(\ref{formalsol}) has to be considered in the formal sense as the convergence of the integral is not guaranteed for arbitrary $\tilde{g}(v)$, and arbitrary value of the parameters $\sigma$, $r_{2}$ and $r_{3}$.

We require that at $t=0$ (i.e. $T \to \infty$ and $P \to \infty$ such that the ration $P/T$ is finite) the equation of state for the extended model coincides with the corresponding limit of the standard van der Waals equation of state reducing to a perfect gas of rigid spherical molecules. Consequently,  if the temperature is sufficiently high, possible thermodynamic effects that could be described by the parameter $r_{2}$ and $r_{3}$ are assumed to be negligible as well as the electromagnetic interaction encoded in the mean field parameter $a$.

\noindent Under this condition, a direct comparison between the mean field partition function obtained in \cite{BM} suggests to choose
\[
\sigma = -R \qquad \tilde{g}(v) = R  \log(v-b).
\]
It is straightforward to verify that, with this choice, the solution of the form~(\ref{formalsol})  at $t=0$, and in the limit $\eta \to 0$, reduces to the van der Waals equation of state.

Observing that exponent in~(\ref{formalsol}) behaves as
\[
-\frac{x v + \vep_{0}{v} t - R  \log(v-b)}{ R \eta} \simeq -\frac{(x-r_{2} t) v}{R \eta}
\] 
in the limit $v \to \infty$, the convergence of the integral~(\ref{formalsol}) for all $x>0$ and $t>0$ is guaranteed by the request
\[
r_{2} \leq 0.
\]
Finally, the required solution to the equation~(\ref{phieq2}) is
\begin{equation}
\label{phisol}
\varphi(x,t) = \int_{b}^{\infty} \; \exp \left({-\frac{x v + \vep_{0}(v) t - R \log(v-b) }{R \eta}} \right) \; dv,
\end{equation}
naturally interpreted as the mean field partition function for the IEW model.
Evaluating the molar volume at $t=0$, using the formula~(\ref{colehopf}), we have
\[
v(x,0) = - R \eta \;  \der{\log \varphi(x,0)}{x} = b + \frac{R}{x} + \frac{R \eta}{x}
\]
that, in the limit $\eta \to 0$, coincides with the van der Waals equation evaluated at $t=0$.

\section{ Critical asymptotics and phase diagram}
\label{sezione 4}
As discussed in sections above, our model assumes the formal expansion~(\ref{epsexp}) and its solution is given by the partition function~(\ref{phisol}). Due to the small parameter $\eta$, thermodynamic properties of the model can be determined via the asymptotic evaluation of the partition function~(\ref{phisol}).

For convenience, let us introduce the function
\[
\Phi(x,t,v) = \frac{1}{R} \,\big[ x v + \vep_{0} t - R \log (v-b) \big] .
\]
Using the standard Laplace's formula, the partition function~(\ref{phisol}) can be approximated, at the leading order, as follows
\begin{equation}
\label{asymsol}
\varphi(x,t; \eta) = \int_{b}^{\infty} \; e^{-\Phi/\eta} \; dv \simeq   \sum_{k} \sqrt{\frac{2 \pi \eta}{  \Phi_{k}''  }} \; \mathrm{e}^{-\Phi_k/\eta  } \, 
\end{equation}
where the sum index runs over the local minima of the function $\Phi(x,t,v)$ at fixed $x$ and $t$, $\Phi' = \partial \Phi / \partial v$, $\Phi'' = \partial^{2} \Phi / \partial v^{2}$ and $\Phi_{k} \equiv \Phi_{k}(x,t) = \Phi(x,t,v_{k}(x,t))$ where $v_{k}(x,t)$ is a solution of the equation
\begin{equation}
\label{phieqs}
\Phi'(x,t,v_{k}) = 0,
\end{equation}
that is equivalent to the equation of state~(\ref{eqstate_part}). Then, $v_{k}$'s identically satisfy
\[
x  + \left [ \frac{r_{2} v_{k}^{2} + a}{(v_{k}- R r_{3})^{2}} - \frac{2 r_{2} v_{k}}{v_{k}- R r_{3}} \right] t - \frac{R}{v_{k}- b} = 0.
\]
We point out that the function $\Phi(x,t,v)$ is related to the Gibbs free energy up to a factor $1/t$ and an additive function of the variable $t$ only and therefore the above critical points provide the equation of state for the thermodynamic system.

\begin{figure}[h]
\centering
\includegraphics[width=3.8cm]{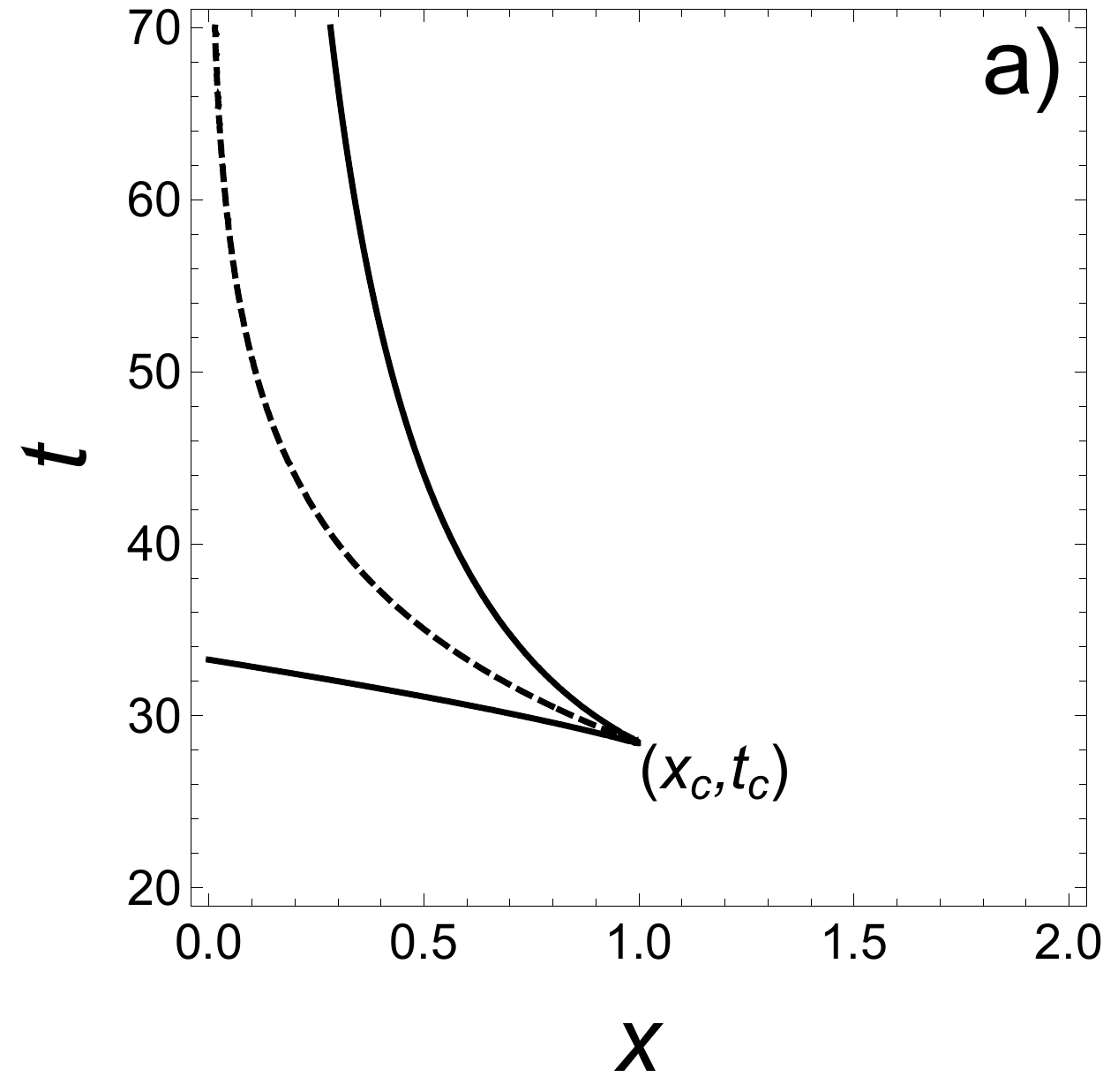} 
\hspace{0.4 cm} \includegraphics[width=3.8cm]{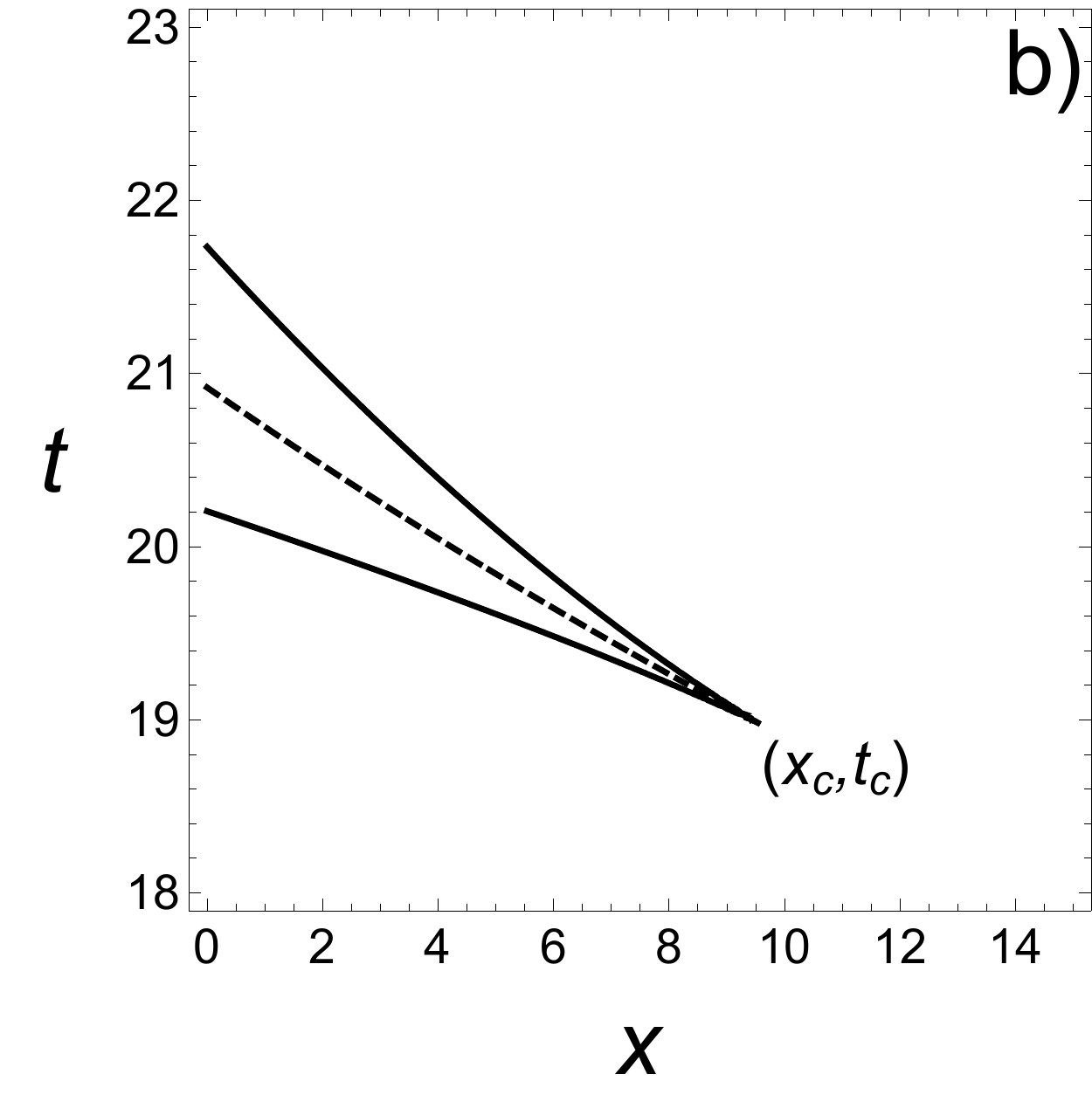}
\caption{$a$) Phase diagrams for the van der Waals model $b$)  and the choice (\ref{cond2a}) with $R \simeq 8.3145\; J/mol K$, $a=1 \; Pa \, m^6 \, mol^{-2}$, $b=1 \; m^3 \, mol^{-1}$.}
\label{phasext}
\end{figure}

The critical point $(x_{c},t_{c},v_{c})$ is obtained as a simultaneous solution to the equations
\[
\Phi(x_{c},t_{c},v_{c}) = 0 \qquad \Phi'(x_{c},t_{c},v_{c}) = 0 \qquad \Phi''(x_{c},t_{c},v_{c}) = 0.
\]
Solving the above relations we obtain
\begin{gather}
\label{eqs punti critici}
\begin{aligned}
&v_{c} = 3 b - 2 R r_{3} \qquad x_{c} =\frac{R}{8(b- R r_{3})} + \frac{27 R r_{2} (b- R r_{3})}{8 (a + R^{2} r_{2} r_{3}^{2})} \\
&t_{c} = \frac{27 R (b-R r_{3})}{8 (a + R^{2} r_{2} r_{3}^{2})}.
\end{aligned}
\end{gather}
The extended model parametrised by $r_{2}$ and $r_{3}$ admits a physical critical point, provided  that formulae (\ref{eqs punti critici}) do not conflict with the fundamental inequalities
\[
v_{c} > b \qquad x_{c} >0 \qquad t_{c} > 0,
\]
In virtue of this, and bearing in mind that $r_2$ is restricted in sign, we conclude that admissible critical points do exist if
\begin{align}
\label{rcond1}
r_{3} < \frac{b}{R} \qquad -\frac{a}{R^{2} r_{3}^{2} + 27 (b-R r_{3})^{2}} < r_{2} \leq 0.
\end{align}
In Figure~\ref{phasext}, for illustrative purposes, we compare the phase diagram for the van der Waals model $r_{2}=0$, $r_{3}=0$ and the similar diagram for the extended model with the particular choice
\begin{equation}
\label{cond2a}
r_{2} =- \frac{55}{54} \frac{a}{b^{2}}  \qquad r_{3} =\frac{27}{28} \frac{b}{R}.
\end{equation}
The two solid lines intersect at the critical point $(x_{c},t_{c})$ and identify the {\it critical region}, that corresponds to the value of thermodynamic parameters for which  the equation of state admits three distinct roots. Two roots are associated to the local minima of the free energy function $\Phi$ that are stable equilibrium states of the system; the remaining one is a local maximum associated to an unstable state. The coexistence line (dashed line in Figure~\ref{phasext}) corresponds to the set of points where the free energy admits two minima of equal magnitude, and it is interpreted as the state where two phases (gas and liquid) coexist. Combining the formulae~(\ref{colehopf}) and~(\ref{asymsol}),  we obtain the following asymptotic expression for the molar volume
\begin{equation}
v \simeq R \;  \der{\Phi_{l}}{x} \qquad \eta \to 0
\end{equation}
where $\Phi_{l} = \Phi(x,t,v_{l}(x,t))$ is evaluated on the root $v_{l}(x,t)$ of the equation of state~(\ref{phieqs}) where the free energy $\Phi$ attains the lowest minimum.
Hence, the molar volume $v$ plays the role of order parameter and develops a discontinuity along the coexistence curve associated to a first order phase transition.

\section{Comparison with empirical model equations}
\label{sezione 5}
This section is aimed at a comparison between the equation of state~(\ref{eqstate_part}), that  results from the asymptotic analysis of the IEW model, and two of the most popular semi-empirical equations of state introduced in the chemical engineering literature: Soave's modification of the Redlich-Kwong equation \cite{Soave,RK} and the Peng-Robinson equation \cite{PR}. 
This comparison will provide some interesting insights on the generality, actual potential and effectiveness of the extension procedure discussed above.

We look for suitable choices of the parameters $r_2$ and $r_3$  such that the critical point of the IEW model coincides with the one of SRK and PR models and compare both critical and off critical isotherms. For illustrative purpose, van der Waals parameters are chosen for the hydrogen gas.

For convenience, let us introduce the following parametrisation of the constants $r_2$ and $r_3$
\begin{equation*}
r_2= k_2 \frac{ \, a}{b^2} \qquad \qquad r_3=k_3 \frac{ \, b}{ R} \, ,
\end{equation*}
where $k_2$ and $k_3$ are (in general) two dimensionless functions of the van der Waals parameters $a$ and $b$. Hence, the equation of state in the original variables $v$, $T$ and $P$ reads as follows
\begin{equation}
\label{genvdWPT}
P= \frac{R T}{v-b} - \frac{a}{\left( v-  k_3  \, b \right)^2} \left[ 1+2k_2 k_3 \frac{v}{b} -k_2 
\left( \frac{v}{b} \right)^2 \right] \, .
\end{equation}	
	
\subsection{Soave-Redlich-Kwong equation.}	
In \cite{Soave}, Soave proposed the following refined version of the  equation of state previously introduced by Redlich-Kwong \cite{RK}
\begin{equation}
\label{SRKeq}
P= \frac{R T}{v-b} - \frac{a(T)}{v(v+b)} \, ,
\end{equation}
where $b$ is the hard core volume and $a(T)$ models attractive forces van der Waals forces between the molecules. Denoting with $T_c$ the critical temperature of the model, we have
\begin{equation*}
a(T)=a_c \left[ 1+(0.480+1.574 \omega -0.176 \omega ^2)(1-\sqrt{T/T_c})\right]^2 \, ,
\end{equation*}
where $a_c=a(T_c)$ and $\omega$ is the so called {\it acentric factor}. 
The acentric factor, introduced by K. S. Pitzer et al. in \cite{pitzer}, takes into account the deviations from sphericity of molecules and it is typically obtained via empirical analysis. 
	
The critical point ($T_{c}, P_{c},v_{c}$) of the SRK model is given by
\begin{equation*}
T_c = 0.2027 \frac{a_c}{b R} \qquad 
P_c = 0.01756\frac{a_c}{b^2} \qquad
v_c = 3.847\, b \, .
\end{equation*}		
One can verify that for $a_c = 1.027 a$ and for the special choice of $k_{1}$ and $k_{2}$
\[
k_2=-0.2417 \times 10^{-3} \qquad k_3=-0.4237
\] 
the critical point of the equation~(\ref{genvdWPT}) coincides with the critical point of the SRK equation (\ref{SRKeq}).


\begin{figure}
\centering
\includegraphics[width=6cm]{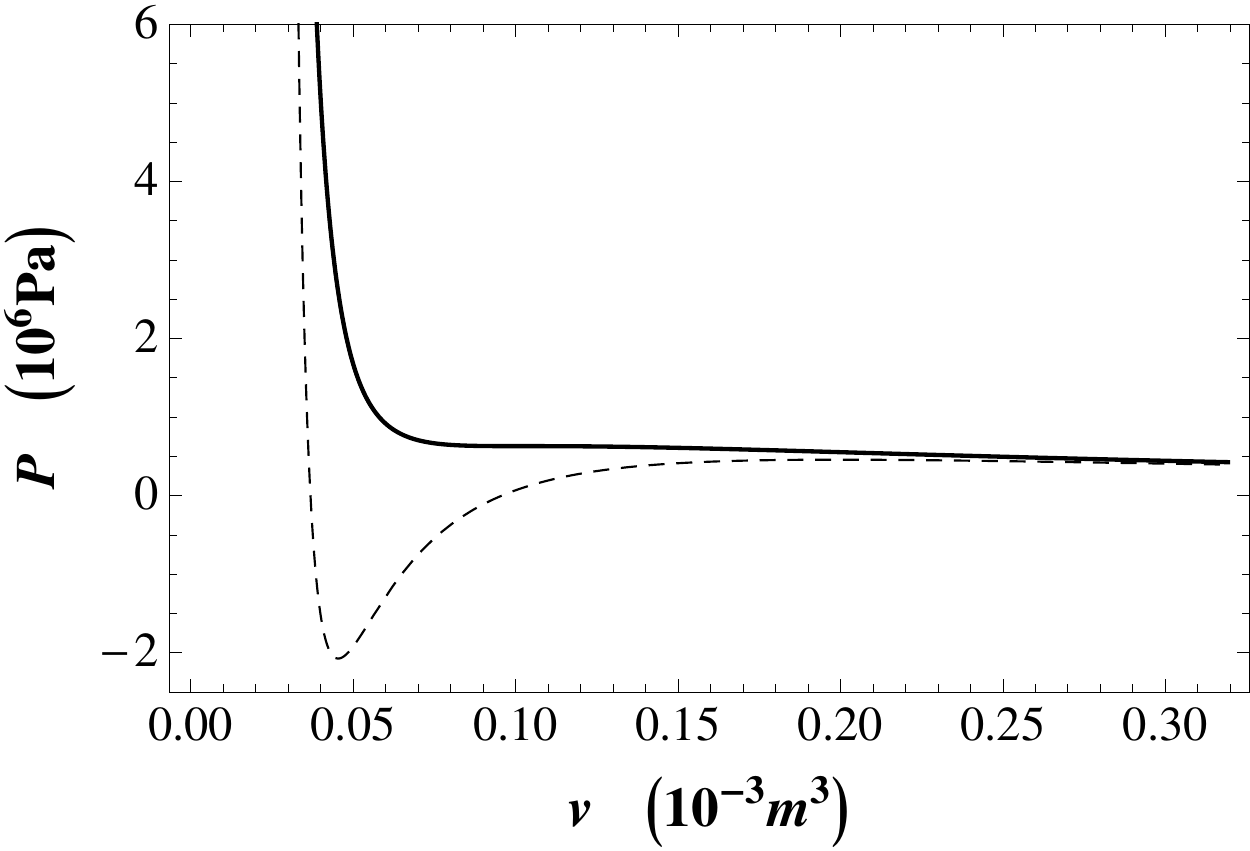}
\caption{Isothermal curves for the hydrogen ($a=2.476 \,10^{-2} Pa \, m^6 \, mol^{-2}$ and $b=2.661 \, 10^{-5} \, m^3 \, mol^{-1}$) at the SRK 
 critical temperature, $T_c =23.29 K$. The dashed line is the van der Waals equation, the thin solid
 line associated to the SRK equation (\ref{SRKeq}), and the thick solid line associated to the IEW equation, with $k_2=-0.2417 \times 10^{-3}$, $k_3=-0.4237$ and $a_c = 1.027 a$, eq. (\ref{genvdWPT}) show an almost perfect overlap.}
\label{FigCritGenSRK}	
\end{figure}
    
\begin{figure}
\centering
\includegraphics[width=6cm]{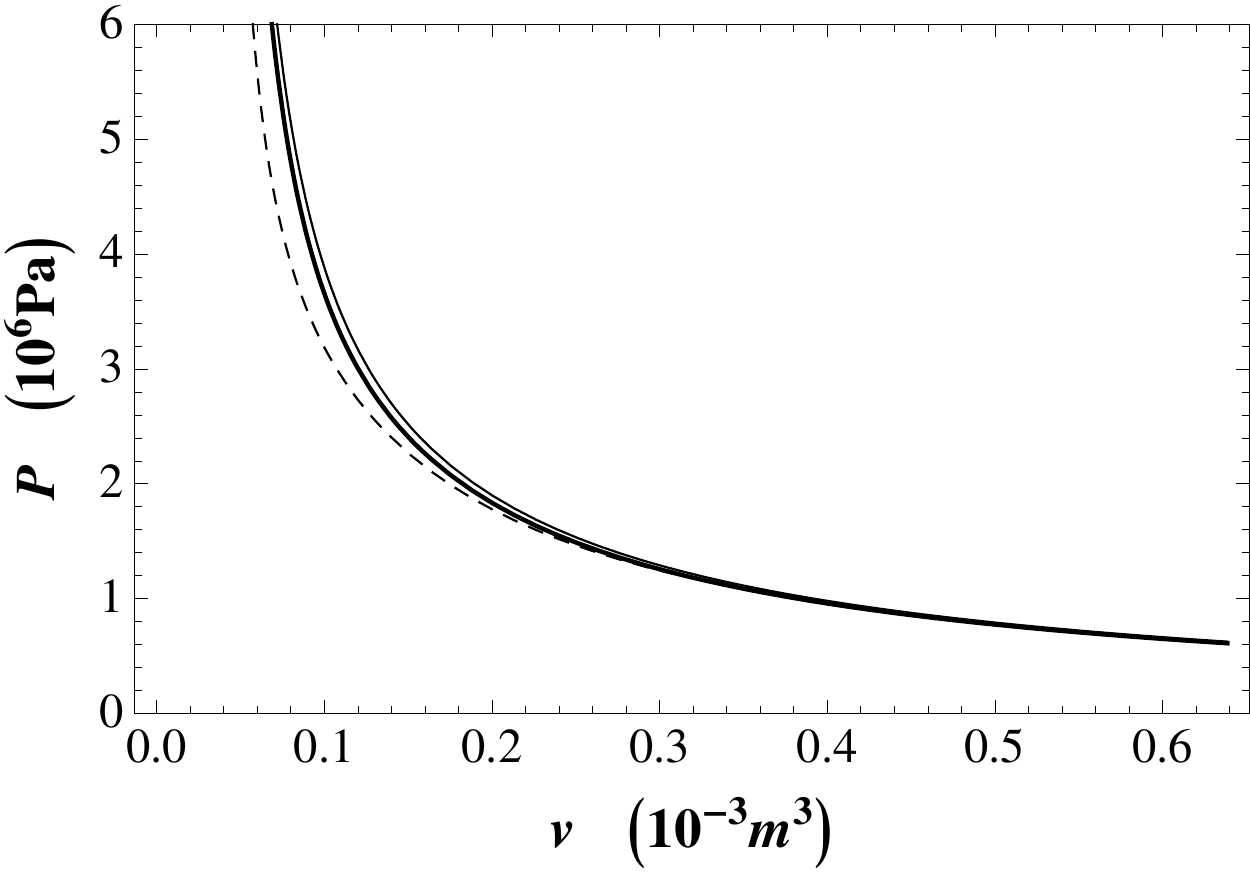}
\caption{Isothermal curves above the critical temperature ($T=50 \, K$) for equations (\ref{SRKeq}) (thin solid line), (\ref{genvdWPT}) with $k_2=-0.2417 \times^{-3}$, $k_3=-0.4237$ and $a_c = 1.027 a$ (thick solid line) and van der Waals isothermal curve (dashed line) for the hydrogen, $a=2.476 \,10^{-2} Pa \, m^6 \, mol^{-2}$ and $b=2.661 \, 10^{-5} \, m^3 \, mol^{-1}.$ }
\label{FigAboveGenSRK}	
\end{figure}

\begin{figure}[h]
\centering
\includegraphics[width=6cm]{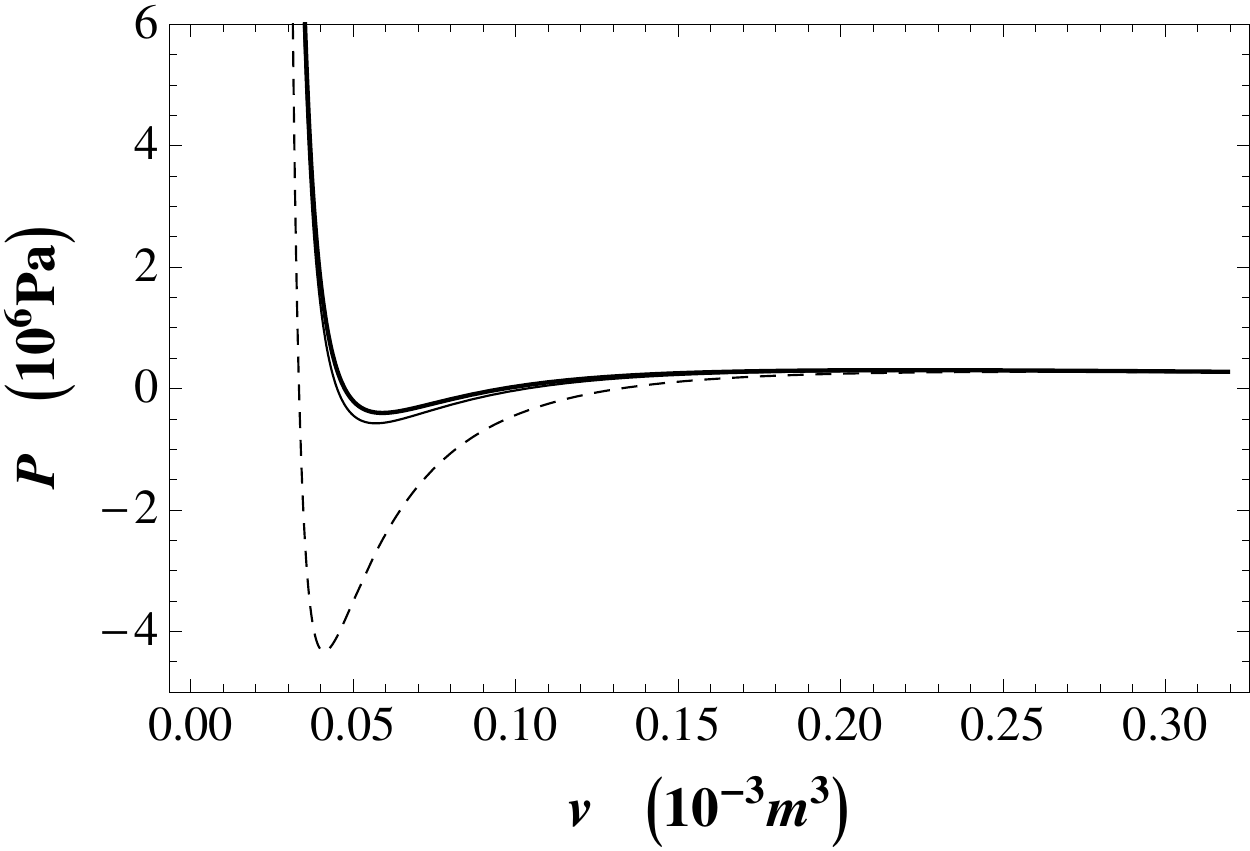}
\caption{Isothermal curves below the critical temperature ($T=18 \, K$) for equations (\ref{SRKeq}) (thin solid line), (\ref{genvdWPT}) with $k_2=-0.2417 \times 10^{-3}$, $k_3=-0.4237$ and $a_c = 1.027 a$ (thick solid line) and van der Waals isothermal curve (dashed line) for the hydrogen, $a=2.476 \,10^{-2} Pa \, m^6 \, mol^{-2}$ and $b=2.661 \, 10^{-5} \, m^3 \, mol^{-1}.$}
\label{FigBelowGenSRK}	
\end{figure}

Figure~\ref{FigCritGenSRK} shows a remarkable overlap between the hydrogen critical isothermal curves of the SRK equation and those of the equation (\ref{genvdWPT}) with the above choice of parameters $k_2$ and $k_3$. The corresponding van der Waals isothermal curve is also reported for further clarity. Figures~\ref{FigAboveGenSRK} and~\ref{FigBelowGenSRK} show  a very good agreement between isothermal curves of both models above and below the critical temperature. The isothermal curves of the SRK equation are considered for the hydrogen acentric factor $\omega = -0.22$ \cite{Yaws}. 
          
\subsection{Peng-Robinson equation.}
\begin{figure}
\centering
\includegraphics[width=6cm]{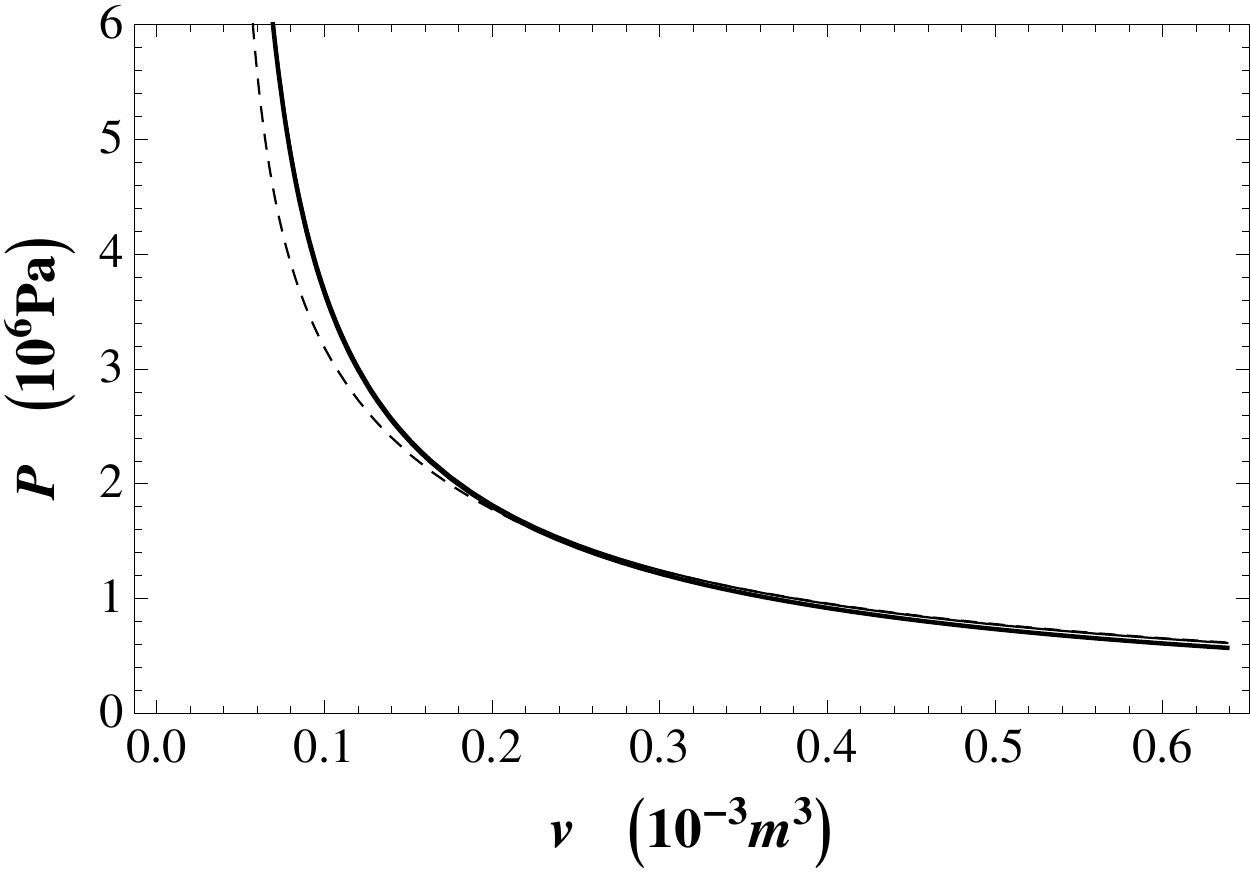}
\caption{Isothermal curves above the critical temperature ($T=50 \, K$) for equations (\ref{PReq}) (thin solid line), (\ref{genvdWPT}) with $k_2=-0.1387 \times 10^{-2}$, $k_3=-0.4757$ and $a_c = 1.180 a$ (thick solid line) and van der Waals isothermal curve (dashed line) for the hydrogen, $a=2.476 \,10^{-2} Pa \, m^6 \, mol^{-2}$ and $b=2.661 \, 10^{-5} \, m^3 \, mol^{-1}.$}
\label{FigAboveGenPR}	
\end{figure}

\begin{figure}[h]
	\centering
	\includegraphics[width=6cm]{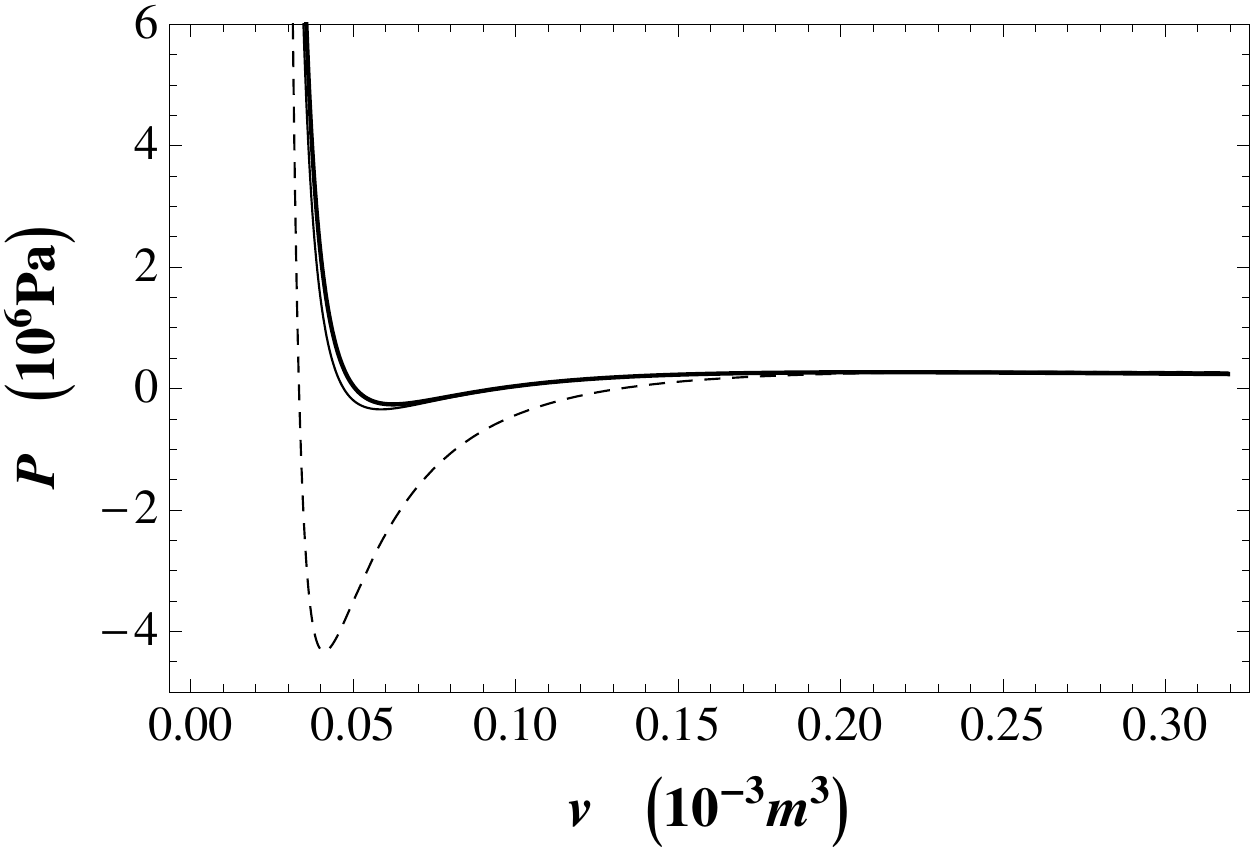}
            \caption{Isothermal curves below the critical temperature ($T=18 \, K$) for equations (\ref{PReq}) (thin solid line),(\ref{genvdWPT}) with $k_2=-0.1387 \times 10^{-2}$, $k_3=-0.4757$ and $a_c = 1.180 a$ (thick solid line) and van der Waals isothermal curve (dashed line) for the hydrogen, $a=2.476 \,10^{-2} Pa \, m^6 \, mol^{-2}$ and $b=2.661 \, 10^{-5} \, m^3 \, mol^{-1}.$}
\label{FigBelowGenPR}	
     \end{figure}

In \cite{PR} Peng and Robinson  proposed the   equation of state	 
\begin{equation}
\label{PReq}
P= \frac{R T}{v-b} - \frac{a(T)}{v(v+b)+b(v-b)} \, , 
\end{equation}
where 
\begin{equation*}
a(T)=a_c \left( 1+(0.3746+1.542 \, \omega -0.2699 \, \omega ^2)(1-\sqrt{T/T_c})\right)^2 \, ,
\end{equation*}
$a_c=a(T_c)$ and $\omega$ is the acentric factor as introduced for the SRK equation of state.
The critical point is 
\begin{equation*}
T_c = 0.1701 \frac{a_c}{b R} \qquad
P_c = 0.01324 \frac{a_c}{b^2} \qquad
v_c = 3.951 \, b \, .
\end{equation*} 
Similarly to the case of the SRK equation, a comparison can be made between our extended model   (\ref{genvdWPT}), and the PR model. 
More precisely, for $a_c = 1.180\, a$ and for the choice
 \[
 k_2=-0.1387 \times10^{-2} \qquad k_3=-0.4757
 \] the critical point of the equation (\ref{genvdWPT}) equals the critical point of the Peng-Robinson equation (\ref{PReq}). 
\begin{figure}
 \centering
  \includegraphics[width=6cm]{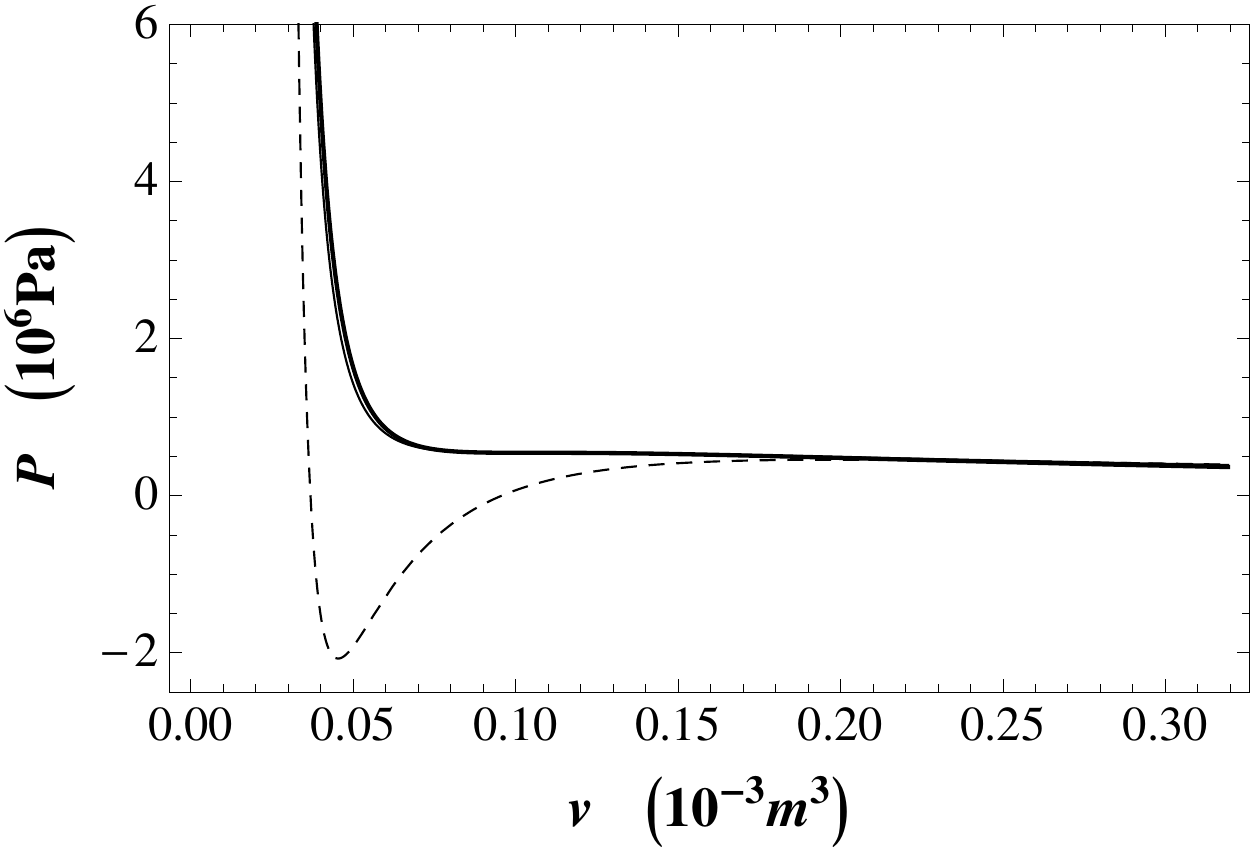}
  \caption{Isothermal curves at the PR critical temperature ($T_c = 22.46 \, K$) for equations (\ref{PReq}) (thin solid line), (\ref{genvdWPT}) with $k_2=-0.1387 \times 10^{-2}$, $k_3=-0.4757$ and $a_c = 1.180 a$ (thick solid line) and van der Waals isothermal curve (dashed line) for the hydrogen, $a=2.476 \,10^{-2} Pa \, m^6 \, mol^{-2}$ and $b=2.661 \, 10^{-5} \, m^3 \, mol^{-1}$.
}
\label{FigCritGenPR}
\end{figure}
Figure~\ref{FigCritGenPR} shows a remarkable overlap for critical isotherms given by the equation (\ref{genvdWPT}) with the corresponding PR critical isothermal curves.
Figures~\ref{FigAboveGenPR} and~\ref{FigBelowGenPR} show isothermal curves respectively above 
and below the critical temperature with the hydrogen acentric factor $\omega=-0.22$.  As for the SRK equation the agreement is very good both above and below the critical temperature. The result of this comparison suggests that the IEW model can be proposed as a single two parameter van der Waals extension that allows to interpolate between the SRK and PR equations of state.
       
\section{Concluding remarks}
\label{sezione 6}
We have introduced a new procedure for the construction of multi-parameter extensions of the standard van der Waals model. Our method relies on the macroscopic formulation of the van der Waals equation of state as a solution to an integrable nonlinear conservation law. Unlike the numerous empirical extensions based on a phenomenological modification of the form of the equation of state, we start from the asymptotic expansion of the internal energy in the small parameter given by the inverse of Avogadro's number and require that the nonlinear PDEs for the molar volume, equivalent to the first law of thermodynamics, is linearisable via a Cole-Hopf transformation. The model so constructed is also required to reproduce the standard van der Waals equation of state in regime of high temperature and far from the critical region. The Cole-Hopf transformation provides a natural integrable extension of the mean field partition function obtained in~\cite{BM} which satisfies a two parameter deformation of the Klein-Gordon equation. The integrability condition implies that similarly to the van der Waals case, there exist infinitely many conservation laws associated to infinitely many state functions and conjugated variables. Such a family of admissible integrable extended models is completely specified by four real constants. Two of these constants can be identified with the standard van der Waals parameters $a$ and $b$ 
d respectively to the electromagnetic mean field interaction and the hard core volume. We have also shown that the two additional deformation parameters can suitably be chosen to match the critical point of the PR and SRK equations of state which have been previously introduced in the literature to provide a more accurate description of the critical properties of solutions and multi-phase systems. With these choices of parameters, isothermal curves of our model are in good  agreement with both PR and SRK isotherms suggesting that the IEW equation of state~(\ref{eqstate_part}) can be used as interpolating model. The procedure presently discussed demonstrates how new models of phenomenological and theoretical relevance can be introduced starting from a simpler model, possibly obtained from first principles, introducing the condition that the extension preserves some key features of the original model, e.g. $C-$integrability~\cite{Calogero}. Integrability implies the existence of infinitely many functions of state and leads to the explicit evaluation of phase diagrams and asymptotic formulae for the isotherms. This suggests that new models could be constructed based on alternative notions of integrability and might have a physical counterpart in the context of complex and multiphase thermodynamic systems. For example, considering higher order terms in the expansion of the free energy of the form~(\ref{Eexp})
\begin{align*}
\vep =& \vep_{0} + \eta \left( \vep_{11} \der{v}{x} + \vep_{12} \der{v}{t} \right) + \eta^{2} \left(\vep_{21} \dersec{v}{x} + \vep_{22} \frac{\partial^{2} v}{\partial x \partial t} + \vep_{23} \dersec{v}{t}\right)  \\
& \dots + g(t)
\end{align*}
where $\vep_{ij} = \vep_{ij}(v)$ the associated equation~(\ref{balance}) can be analysed by using more general approaches to integrability such as e.g. symmetry and perturbative symmetry approaches~\cite{Mik1,Mik2,Mik3}, perturbative approaches to integrable or quasi-integrable (bi-)Hamiltonian conservation laws~\cite{DZ,Dubrovin}, perturbative approaches to non-Hamiltonian dispersive and viscous conservation laws~\cite{ALM,ALM2}. The asymptotic behaviour and the shock structure of solutions to nonlinear conservation laws in the limit of small viscosity-dispersion is as rich as intriguing and could possibly lead to new type of phase transition in thermodynamics of multicomponent and complex systems associated to non-classical and dispersive shocks (see e.g.~\cite{HES}).
 This interesting direction is currently under study and will be the subject of a future publication.

\vspace{.5cm}
\noindent {\bf Acknowledgements.} Authors are grateful to Adriano Barra for useful discussions and numerous precious insights on the theory of phase transitions and collective phenomena. Authors are also indebted to Paolo Lorenzoni for his comments on the first version of the manuscript and useful discussions on integrable conservation laws. F.G. has been partially supported  by the Department of Mathematics and Physics {\it "Ennio de Giorgi"} of the University of Salento through a special young student training program. G.L has been partially supported by 
PRIN 2010 - MIUR "Teorie geometriche e analitiche dei sistemi Hamiltoniani in dimensioni finite e infinite". A.M. has been partially supported by Progetto giovani GNFM-INdAM 2014 "Aspetti geometrici e analitici dei sistemi integrabili". 





\end{document}